\documentclass[12pt,a4paper,oneside]{article}
\usepackage[top=1in,bottom=1in,left=.5in,right=.5in]{geometry}
\usepackage[margin=.5in,small]{caption}

\usepackage{authblk}
\usepackage{cite}
\usepackage{hyperref}
\usepackage{doi}
\usepackage{url}
\usepackage{amsmath,mathtools,gensymb,bigints}
\usepackage{epsfig}
\usepackage{graphics}
\usepackage{amsfonts, amssymb, upgreek}
\usepackage{textcomp}
\usepackage[english]{babel}
\usepackage{blindtext}
\usepackage{color}
\usepackage{multirow}

\hypersetup{
  pdfinfo={
  pdfproducer={},
  Title={KPZ models: height-gradient fluctuations and the tilt method},
  Subject={},
  Author={M. F. Torres and R. C. Buceta},
  Keywords={growth processes: interfaces in random media; kinetic roughening;
classical Monte Carlo simulations},
  Subject={preprint to submit},
  }
}

\providecommand{\abs}[1]{\vert #1\vert}

\newcommand{\ex}{\mathop{}\!\mathrm{e}}

\usepackage[utf8]{inputenc}

\newcommand\keywords[1]%
  {\begin{flushleft}
   \let\and\\%
   \textbf{Keywords:}\\
   #1
   \end{flushleft}%
  }
\begin{document}

\title{KPZ models: height-gradient fluctuations and the tilt method }

\author[1,2]{M. F. Torres\thanks{mtorres@ifimar-conicet.gob.ar}}
\author[1,2]{R. C. Buceta\thanks{rbuceta@mdp.edu.ar}}
\affil[1]{Instituto de Investigaciones F\'{\i}sicas de Mar del Plata, UNMdP and CONICET}
\affil[2]{Departamento de F\'{\i}sica, FCEyN, Universidad Nacional de Mar del Plata}
\affil[{ }]{Funes 3350, B7602AYL Mar del Plata, Argentina}

\maketitle

\abstract{When a growing interface belonging to the KPZ universality class is tilted with average slope $m$, its average velocity increases in  $\frac{\Lambda}{2}\,m^2$, where $\Lambda$ is related to the nonlinear coefficient $\lambda$ of the KPZ equation. Nevertheless, a necessary condition for this association to hold true is that the mean square height-gradient increases in $b\, m^2$ when the interface is tilted. For the continuous KPZ equation $b = 1$ and the relation $\Lambda=\lambda$ is achieved. In this work, we study the local fluctuations of the height gradient through an analysis of the values of $b$. We show that, for 1-dimensional discrete KPZ models, $b$ has a power-law dependence with the discretization step $s$ chosen to calculate the height gradient and $b$ goes to $1$ as $s$ increases. Its power-law exponent $\gamma_b$ matches the exponent associated with the finite-size corrections of the interface average velocity, {\sl i.e.} $\gamma_b=2(\zeta-1)$, where $\zeta$ is the global roughness exponent. We also show how, for restricted (unrestricted) growth models, the value of $b$ goes to $1$ from below (above) as $s$ increases.


\section{Introduction}

The tilt method was introduced by Krug \cite{Krug1989,Krug1990} to show that discrete models of growing interfaces that belong to a universality class can be characterized not only by the exponents and laws of scaling, but also by the nonlinearities present in the system. Usually, to tilt the interface in the simulation of these models, helical boundary conditions are applied \cite{Barabasi1995}, {\sl i.e.} $h_m(L+1) = m L + h_m(1)$, where $h_m$ is the tilted interface height, $L$ is the lateral size, and $m=\langle\nabla h_m\rangle$ is the average slope. Periodic boundary condition corresponds to a non-tilted interface, {\sl i.e.} $m = 0$. Models that  belong to the Kardar-Parisi-Zhang (KPZ) universality class show a dependency between the saturation average velocity and the slope of the tilted interface like \cite{Barabasi1995} 
\begin{equation}
\mathcal{V}_\mathrm{sat}^{(m)}=\mathcal{V}_\mathrm{sat}^{(0)}+\frac{\Lambda}{2}\,m^2\;,\label{eq:v-sal-m}
\end{equation}
for $\abs{m}\ll 1$, where $\Lambda$ is a non-zero real constant and the label $(m)$ refers to an interface with average slope $m$. Otherwise, if $\Lambda=0$ the models are included in the Edwards-Wilkinson (EW) universality class. The linear behaviour of $\mathcal{V}_\mathrm{sat}^{(m)}$ as function of $m^2$ is strictly valid for $\abs{m}\ll 1$; otherwise, other behaviours occur. If the dependency of the average saturation velocity with the slope is different from $\frac{\Lambda}{2}m^2$, the studied model does not belong to the KPZ or EW universality classes.

The quadratic constant $\Lambda$ in equation~\eqref{eq:v-sal-m} is associated with the nonlinear constant $\lambda$ of the KPZ equation \cite{Kardar1986}
\begin{equation}
\frac{\partial h}{\partial t}=\mathcal{F} + \nu\,\nabla^2 h  + \frac{\lambda}{2}\,\abs{\nabla h}^2 + \eta (\mathbf{x},t)\;,\label{eq:KPZ}
\end{equation}
where $h(\mathbf{x},t)$ is the interface height of the $d$-dimensional substratum at the position $\mathbf{x}$ and time $t$. The real constants $\mathcal{F}$, $\nu$ and $\lambda$ describe the growth force, the surface relaxation intensity, and the lateral growth, respectively. The noise $\eta(\mathbf{x},t)$ is Gaussian with zero mean and covariance \mbox{$\langle\eta(\mathbf{x},t)\,\eta(\mathbf{x}',t')\rangle=2\,D\,\delta(\mathbf{x}-\mathbf{x}')\,\delta(t-t')$}, where $D$ is the noise intensity. The average velocity for the tilted KPZ equation~\eqref{eq:KPZ} is
\begin{equation}
\mathcal{V}_\mathrm{KPZ}^{(m)}=\Bigl\langle\frac{\partial h_m}{\partial t}\Bigl\rangle=\mathcal{F} +\frac{\lambda}{2}\bigl\langle\abs{\nabla h_m}^2\bigr\rangle\;,\label{eq:average-v_KPZ-tilted}
\end{equation}
which is a function of the average slope $m$. Notice that the noise average is zero and, if the tilted system is sufficiently large or helical boundary conditions are chosen, the Laplacian average is negligible or zero, respectively. In the saturation, the equation~\eqref{eq:average-v_KPZ-tilted} is independent of time, thus, it is only a function of $m$. The solution for the Fokker-Plank equation associated with the one-dimensional continuous KPZ equation indicates that the local gradients in the interface follow a normal distribution \cite{Halpin1995}. Therefore, tilting the interface only changes the mean value in the gradient distribution. Then, the expansion of the mean square height-gradient (MSHG) is
\begin{equation}
\langle\abs{\nabla h_m}^2\bigr\rangle=\langle\abs{\nabla h_0}^2\bigr\rangle+ m^2+\mathcal{O}(m^4)\;.\label{eq:MSHGKPZ}
\end{equation}
This is the main argument that supports that $\Lambda$ from equation~\eqref{eq:v-sal-m} is equal to $\lambda$. Although for $d>1$ the assumption that the local gradient follows a normal distribution is wrong, it was found that, in models belonging to the KPZ universality class, the equation~\eqref{eq:v-sal-m} still holds \cite{Barabasi1995}.

While the tilt method is a powerful tool to find the nonlinear coefficient $\lambda$ of a given model, there are other possible techniques to obtain it. For example, there are several inverse methods \cite{Lam1993,Campos2013} that search the coefficients used in the discrete integration of the KPZ that best predict the evolution of the interface in the simulation of the KPZ models. Also, $\lambda$ can be obtained from the scaling parameter of the height moments \mbox{$\Gamma=\abs{\lambda}\,A^{1/\zeta}$}, representing the non-stationary fluctuations amplitude of the interface, where $A$ is the amplitude of the stationary height-difference correlation function and $\zeta$ is the global roughness exponent \cite{Krug1992,Alves2016}. These methods usually reproduce a value close to $\Lambda$. For some models, like the single step model, $\lambda$ can be obtained directly from the evolution rules, assuming that the boundary conditions of the tilt method are applied \cite{Barabasi1995}.  

The discrete nature of many of the KPZ models and the finite size of the simulations have an impact in measurements over the interface \cite{Krug1990-2,Kim1991,Alves2014,Alves2016,Reis2004,Daryaei2020,Ferrari2011}. For example, in the saturation of KPZ models, the measured average velocity of a non-tilted interface verify \cite{Krug1990-2}
\begin{equation}
  \mathcal{V}_{\mathrm{sat}}^{(0)}-\mathcal{V}_{\mathrm{sat}}^{(0)}\rfloor_\infty\sim-\lambda\,L^{2(\zeta-1)}\;,
\end{equation}
for $t\gg L^z$, where $L$ is the system size, $\mathcal{V}_{\mathrm{sat}}^{(0)}\rfloor_\infty$ is the velocity value in the thermodynamic limit, and $z$ is dynamic exponent. For the KPZ universality class, the roughness exponent $\zeta<1$. Assuming that the relationship between the average velocity and the mean square height-gradient is given by equation~\eqref{eq:average-v_KPZ-tilted}, derived from the continuous KPZ equation, we obtain
\begin{equation}
\langle\abs{\nabla h_0}^2\bigr\rangle-\langle\abs{\nabla h_0}^2\bigr\rangle_{\!\infty}\sim -L^{2(\zeta-1)}\;,\label{eq:MSHG-vs-L}
\end{equation}
where $\langle\abs{\nabla h_0}^2\bigr\rangle_{\!\infty}$ is the MSHG value in the thermodynamic limit. 

In this work, we focus on the study of local properties of the MSHG in KPZ models. We measure the MSHG by discretizing the height gradient, over the interface, with fix-length step $s$. We show that the MSHG of a tilted interface, for models belonging to the KPZ universality class, verify
\begin{equation}
\langle\abs{\nabla h_m}^2\bigr\rangle=\langle\abs{\nabla h_0}^2\bigr\rangle+ b_s\,m^2\;,\label{eq:MSHG-h_m}
\end{equation}
being the coefficient $b_s$ a function of the discretization step $s$ given by
\begin{equation}
b_{s}=1\pm a_b\,{s}^{\gamma_b}\,,\label{bequa}
\end{equation}
where the minus sign corresponds to a restricted growth model ({\it e.g.} step model), the plus sign to an unrestricted one ({\it e.g.} ballistic deposition model), and $a_b>0\,$.
We also obtain a similar result for the numerical integration of the KPZ equation with restrictions to avoid divergences \cite{Torres-18}. 

\section{Method basics\label{sec:2}}

The height gradient $\nabla h_m$ of a tilted growing interface is subjected to fluctuations that depend on the interface slope and the non-tilted height gradient, which in mean value must be zero, so that \mbox{$\langle\nabla h_m\rangle = m$}. The most obvious proposal for this dependence is 
\begin{equation}
\nabla h_m = m + F(m,\nabla h_0)\;,\label{eq:h_m}
\end{equation}  
where $F$ is a generalized function that indicates the fluctuations of the tilted height gradient and verifies \mbox{$\langle F(m,\nabla h_0)\rangle=0$} and \mbox{$F(0,\nabla h_0)=\nabla h_0$}. A simple calculation allows us to obtain
\begin{equation}
\bigl\langle\abs{\nabla h_m}^2\bigr\rangle = m^2 + \bigl\langle[F(m,\nabla h_0)]^2\bigr\rangle\;,\label{eq:mean-square-grad}
\end{equation}
where the mean square fluctuation (MSF) $\langle F^2\rangle$ is a function of $m$.

To measure the square gradient we took a simple centered discretization \cite{Buceta2005}:
\begin{equation}
\nabla h_m(i,s)=\frac{h_m(i+s)-h_m(i-s)}{2\,s},\label{eq:grad-discr-tilted}
\end{equation}
where $i=1,\dots,L$ is the position in the interface where the gradient is measured and $s$ is a natural number. A larger step $s$ indicates a wider observation window.

For tilted interfaces, it is easy to prove that the height-difference correlation function of second order is
\begin{equation}
G_2^m(2\,s)=\bigl\langle (2s)^2\,[\nabla h_m(i,s)]^2\bigr\rangle\;.
\end{equation}
This correlation, for a non-tilted interface with Family-Vicsek scaling like the KPZ models, behaves as $G_2^0(2\,s)\sim (2\,s)^{2\,\zeta}\,$. Then, the MSHG of a non-tilted interface is
\begin{equation}
\langle[\nabla h_0(i,s)]^2\rangle =\frac{G_2^0(2\,s)}{(2\,s)^2}=a_0\,s^{2(\zeta-1)}\;.\label{G0}
\end{equation}
      
\section{Results for several models and equations\label{sec:3}}

\paragraph*{\bf RSOS model.}
The interface evolution of the \mbox{(1+1)}-dimensional restricted solid-on-solid (RSOS) model is established by the following rule: after choosing a random column, if the height of its two first-neighbouring columns is greater or equal to the chosen one, it grows one unit. Symbolically, if \mbox{$h(i+1)-h(i)\ge 0$} and \mbox{$h(i-1)-h(i)\ge 0$}, where $i$ is the column chosen at random, then \mbox{$h(i)\to h(i)+1$}.

Plot (a) of Figure~\ref{RSOStilt} shows $\langle[\nabla h_0(s)]^2\rangle$ as a function of the discretization step $s$ for several sizes $L$. As expected, they show a power-law behavior with a measured roughness exponent $\zeta$ approaching the theoretical value $0.5$ of the KPZ equation as the system size $L$ increases. The measured values of $\zeta$ are presented in Table~\ref{tabledata}. Plot (b) of Figure~\ref{RSOStilt} shows the linear dependence of the MSHG with the square average slope $m^2$ of the tilted interface. Notice that, as the discretization step $s$ increases, the mean square value of the height-gradient fluctuation becomes less relevant. Plot (c) of Figure~\ref{RSOStilt} shows the calculated value $b_{s}\,$, the coefficient of the linear dependence between  $\bigl\langle[\nabla h_m(s)]^2\bigr\rangle$ and $m^2$, as a function of the step $s$. It is observed that as $s$ approaches 1, the values of $b_s$ move away from 1, an indication of the increase of height-gradient fluctuations. In contrast, large steps $s$ make the values of $b_s\approx 1$. The behavior becomes closer to the one expected for the continuous KPZ and the height-gradient fluctuations diminish. The inset plot (c) shows that $1-b_s$ follows a power-law with $s$, according to equation~\eqref{bequa}. The measured exponent $\gamma_b$ is shown for several system sizes in  Table~\ref{tabledata}.

\begin{figure}[h!]
\centering
\hspace{-2ex}\includegraphics[scale=0.25]{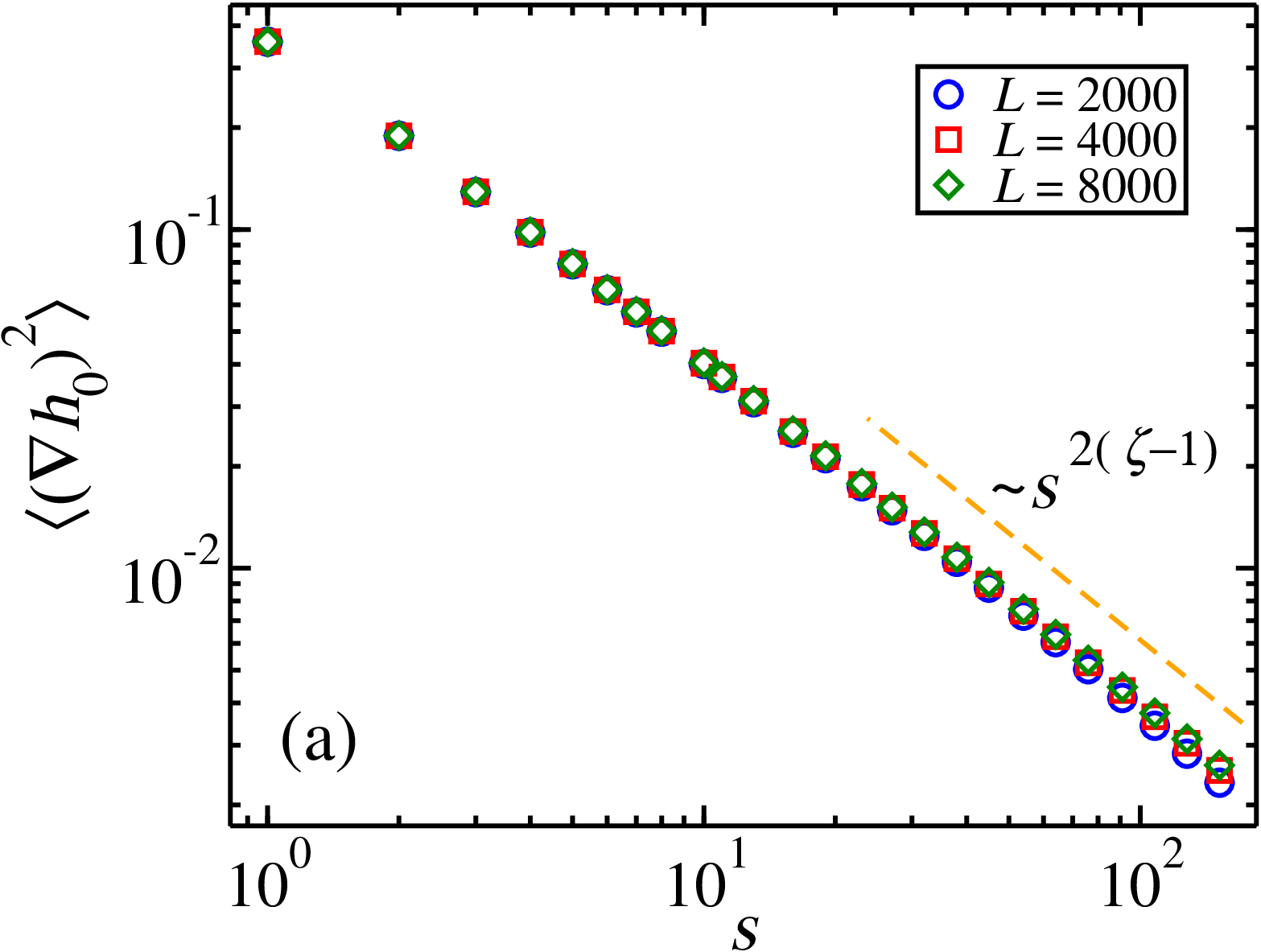}\vspace{3ex}\\
\hspace{-2ex}\includegraphics[scale=0.25]{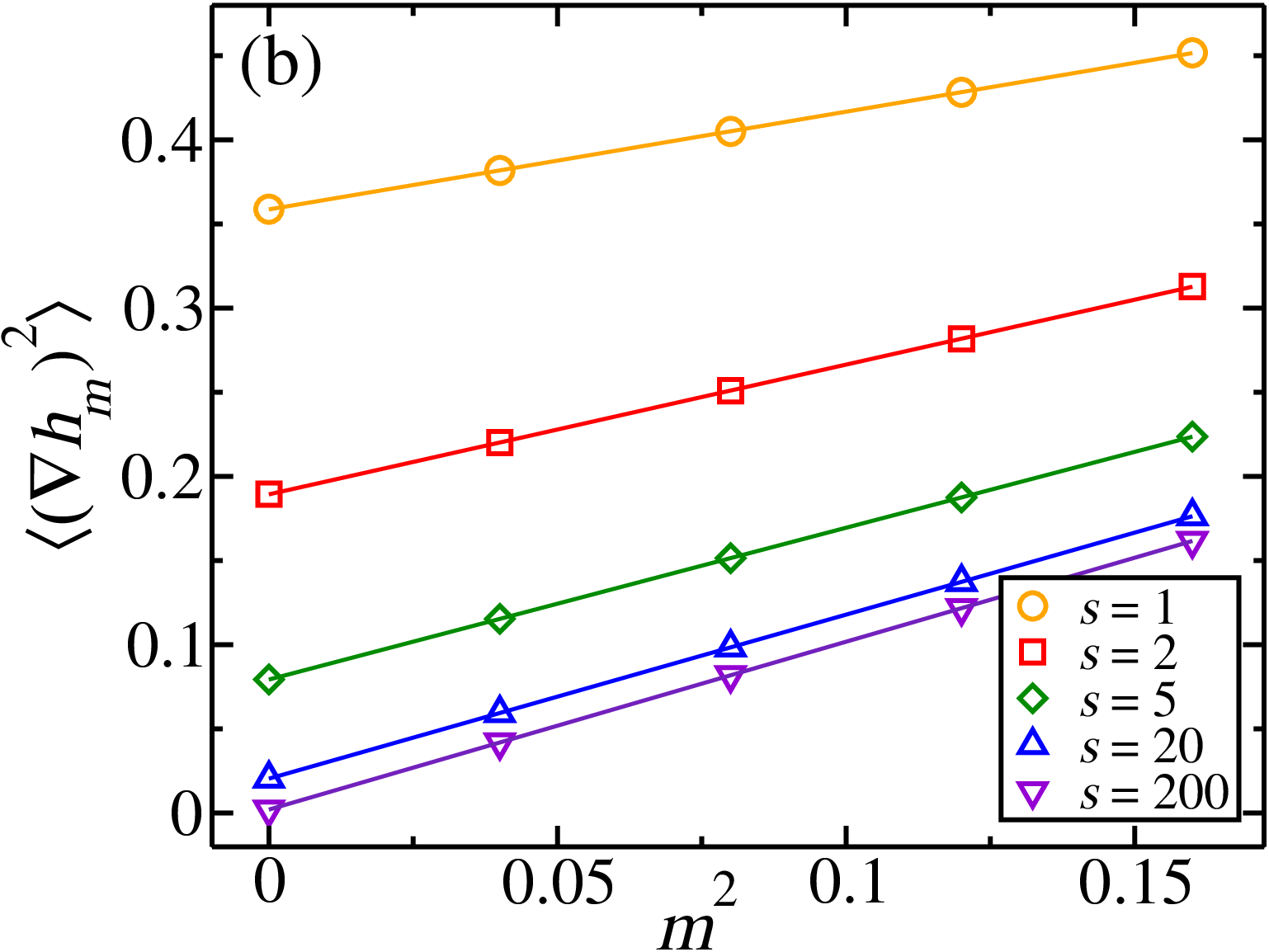}\vspace{3ex}\\
\includegraphics[scale=0.25]{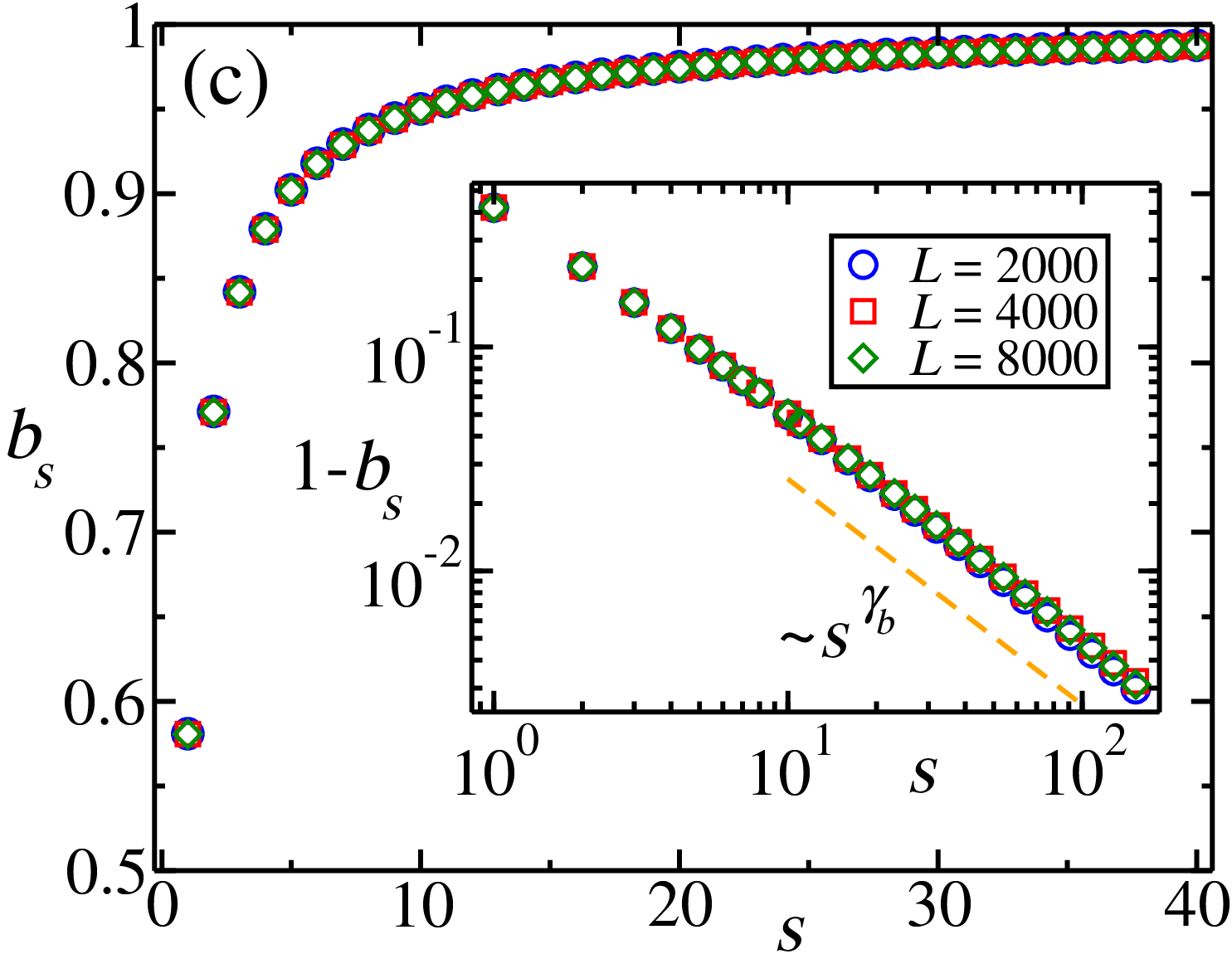}
\caption{{\bf RSOS model.} (a) For a non-tilted interface, MSHG as a function of the discretization step $s$ for several system sizes $L$. The doted-line indicates the power-law dependence. The measured values of $\zeta$, presented in Table~\ref{tabledata}, are close to the theoretical value $0.5$.  (b) For a tilted interface, MSHG as a function of the square average slope $m$ of the interface, for size $L=8000$ and several values of $s$. The colored curves indicate the linear regression. (c) The plot shows how the coefficient $b_s$ as a function of the discretization step $s$ rapidly converges to $1$ from below for several values of size $L$. The inset plot shows how the difference between $1$ and $b_s$ follows a power law as a function of $s$, with the measured exponent $\gamma_b$ presented in Table~\ref{tabledata}.}\label{RSOStilt}
\end{figure}

\paragraph*{\bf Ballistic deposition model.}
The evolution of the \mbox{(1+1)}-dimensional BD model is given by the following rule: the height of the chosen column $h(i)$ grows to $\max[h(i-1),h(i)+1,h(i+1)]$.  The BD model is known for having finite-size dependencies in its exponents due to its unrestricted nature. We repeat the measurements made for the RSOS model, but now for the BD model. The results are plotted in Figure~\ref{BDtilt}. 

The main difference with respect to the RSOS model is that $b_s$ goes to $1$ from above. This is indicated with a plus sign in the equation~\eqref{eq:v-sal-m}. The measured value of $\gamma_b$ presented in Table~\ref{tabledata} is very close to the ones from the RSOS model. 

\begin{figure}[h!]
\centering
\hspace{-2ex}\includegraphics[scale=0.25]{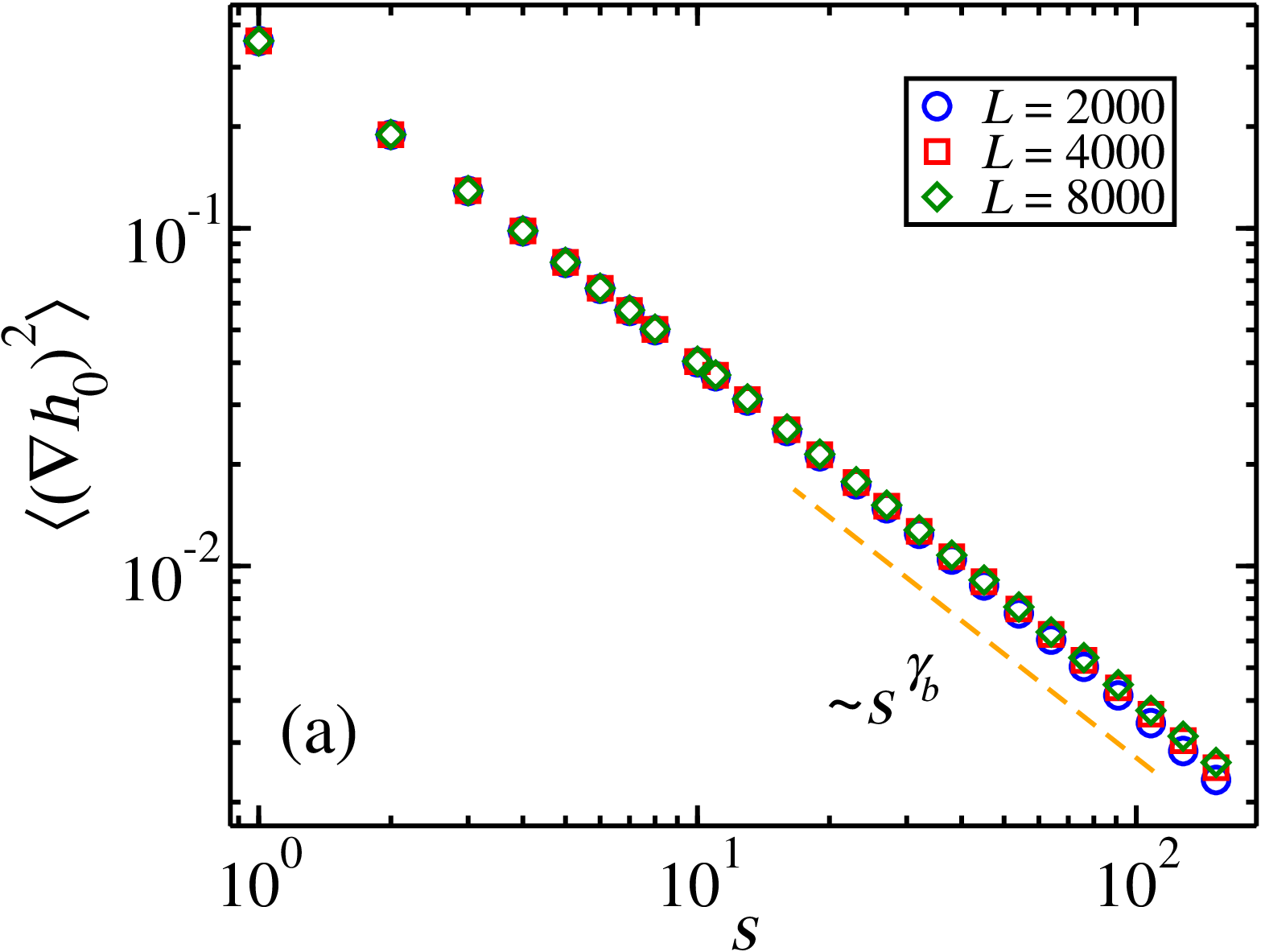}\vspace{3ex}\\
\includegraphics[scale=0.25]{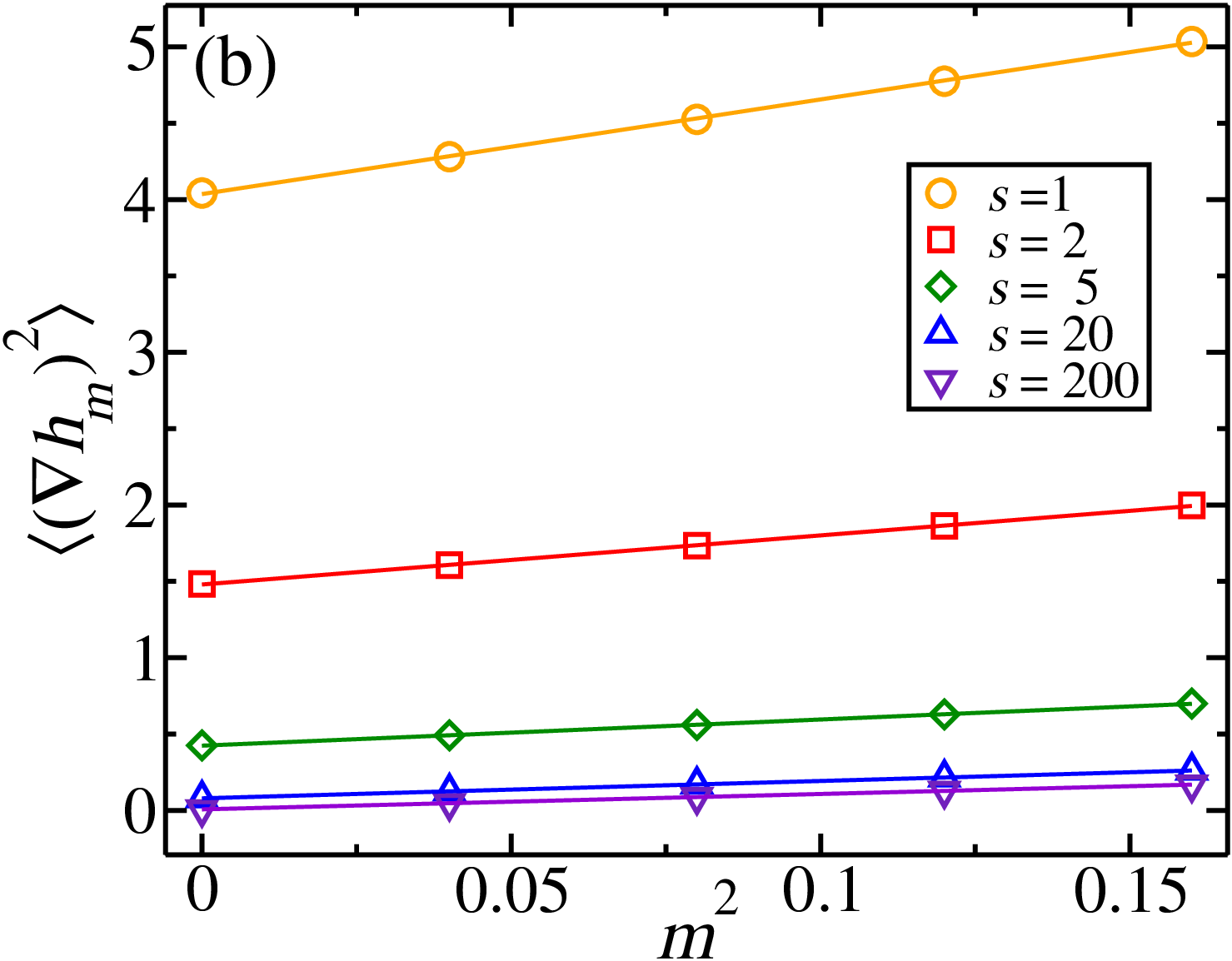}\vspace{3ex}\\
\hspace{1ex}\includegraphics[scale=0.25]{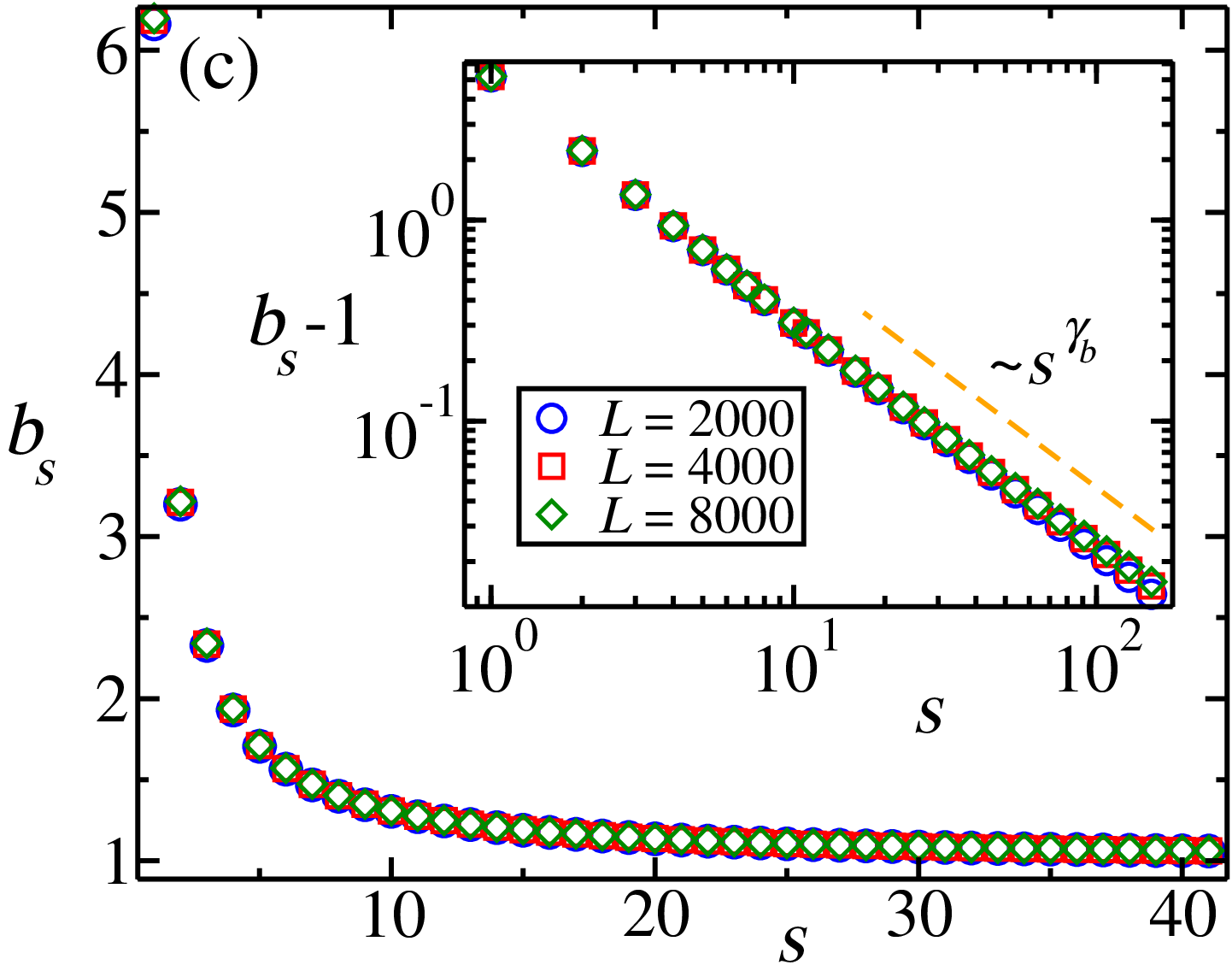}
\caption{{\bf BD model.} (a) For a non-tilted interface, MSHG as a function of the discretization step $s$ for several system sizes $L$. The doted-line indicates the power-law dependence. The measured values of $\zeta$, presented in Table~\ref{tabledata}, are close to the theoretical value $0.5$. (b) For a tilted interface, MSHG as a function of the interface slope $m$ for size $L=8000$ and several values of $s$. The colored curves indicate the linear regression. (c) The plot shows how the coefficient $b_s$ rapidly converges to $1$ from above as a function of the discretization length $s$ for several values of size $L$, in agreement with the equation~\eqref{eq:MSHG-h_m}. The inset plot shows how the difference between $1$ and $b_s$ follows a power law as a function of $s$, with the measured exponent $\gamma_b$ presented in Table~\ref{tabledata}.  }\label{BDtilt}
\end{figure}

\begin{table}[h]
\centering
\begin{tabular}{|c||c|c|c|c|}
\hline
  \multirow{2}{*}{L} & \multicolumn{2}{c|}{RSOS} & \multicolumn{2}{c|}{BD} \\ \cline{2-5} 
                     &  $\zeta$ $(\delta\zeta)$ & $\gamma_b$ $(\delta \gamma_b)$ & $\zeta$ $(\delta\zeta)$ & $\gamma_b$ $(\delta \gamma_b)$ \\ \hline
  $2000$             & $0.468(3)$  & $-1.055(3)$ & $0.438(1)$ & $-1.135(3)$\\ \hline
  $4000$             & $0.486(2)$  & $-1.020(3)$ & $0.458(2)$ & $-1.098(1)$ \\ \hline
  $8000$             & $0.494(1)$  & $-1.036(5)$ & $0.469(2)$ & $-1.081(2)$ \\ \hline 
\end{tabular}
\caption{The second and fourth columns show the measured values of the global roughness exponent $\zeta$ for RSOS and BD respectively. The third and fifth columns show the measured values of the exponent $\gamma_b$ associated with the linear coefficient $b_s$ from equation~\eqref{eq:MSHG-h_m} for RSOS and BD, respectively.  The values inside parentheses are their respective regression errors.}
\label{tabledata}
\end{table}

\paragraph*{\bf Discrete integration of the KPZ equation.}
The numerical integration of the KPZ equation has divergences caused by the uncontrolled growth of pillars (or grooves) over the interface.  To avoid these divergences, some type of modification of the equation is needed. This modification usually implies the replacement of the nonlinear term in the equation for a function $f((\nabla h)^2)$. Dasgupta \textit{et. al.} \cite{Dasgupta1996,Dasgupta1997} successfully take $f((\nabla h)^2)=\frac{1}{c}\bigl(1-\ex^{-c(\nabla h)^2}\bigr)$, where the length $c$ is chosen in a way that avoids the divergences by smoothing the nonlinearities while maintaining all the scaling properties of the equation. Another way is to restrict the value of $(\nabla h)^2$ to a length $\varepsilon$ \cite{Torres-18}, chosen just to eliminate the values of the nonlinearity that trigger the divergence. Here, we use the latter method with a first neighbor discretization of the KPZ equation, for which we replace $(\nabla h_i)^2$ for $f=\min\bigl((h_{i+1}-h_{i-1})^2/4\,,\varepsilon\bigr)$.

For the integration we take $\lambda^2\,D/\nu^3\approx 15\,$, for which the roughness shows the longest power-law behaviour \cite{Moser1991,Torres-18}. In Figure~\ref{Intplot} we show the calculated value of $b_s$ (the linear constant of $\langle[\nabla h_m(s)]^2\rangle$ as a function of $m^2$) as a function of $s$ for several values of $\varepsilon$. Obviously, because we already put a first-neighbor discretization, $b_s$ is already very close to $1$ for $s=1$, but $b_s$ gets closer to $1$ in the same way that the discrete models do. In the inset plot of Figure~\ref{Intplot} we can see that $b_s-1$ follows a power law, like the BD model. This is because the restriction is over the value of the nonlinearity in the integration; unlike the RSOS model, where the restriction is over the interface. The value of $\gamma_b$ measured is presented in Table~\ref{tabledataint}.

As an interesting remark, the measured $\Lambda$ (see Table~\ref{tabledataint}) is not equal to the value $\lambda$ that we input in the integration, although it is close. This result was reached by integrating a modified version of the KPZ equation, not its original version. We recover the input value $\lambda$ with great precision if we take $\Lambda/b_f$, where $b_f$ is the linear coefficient of $\langle f((\nabla h)^2)\rangle$ as a function of $m^2$. These results are presented in Table~\ref{tabledataint}.

\begin{figure}[h!]
\centering
\includegraphics[scale=0.25]{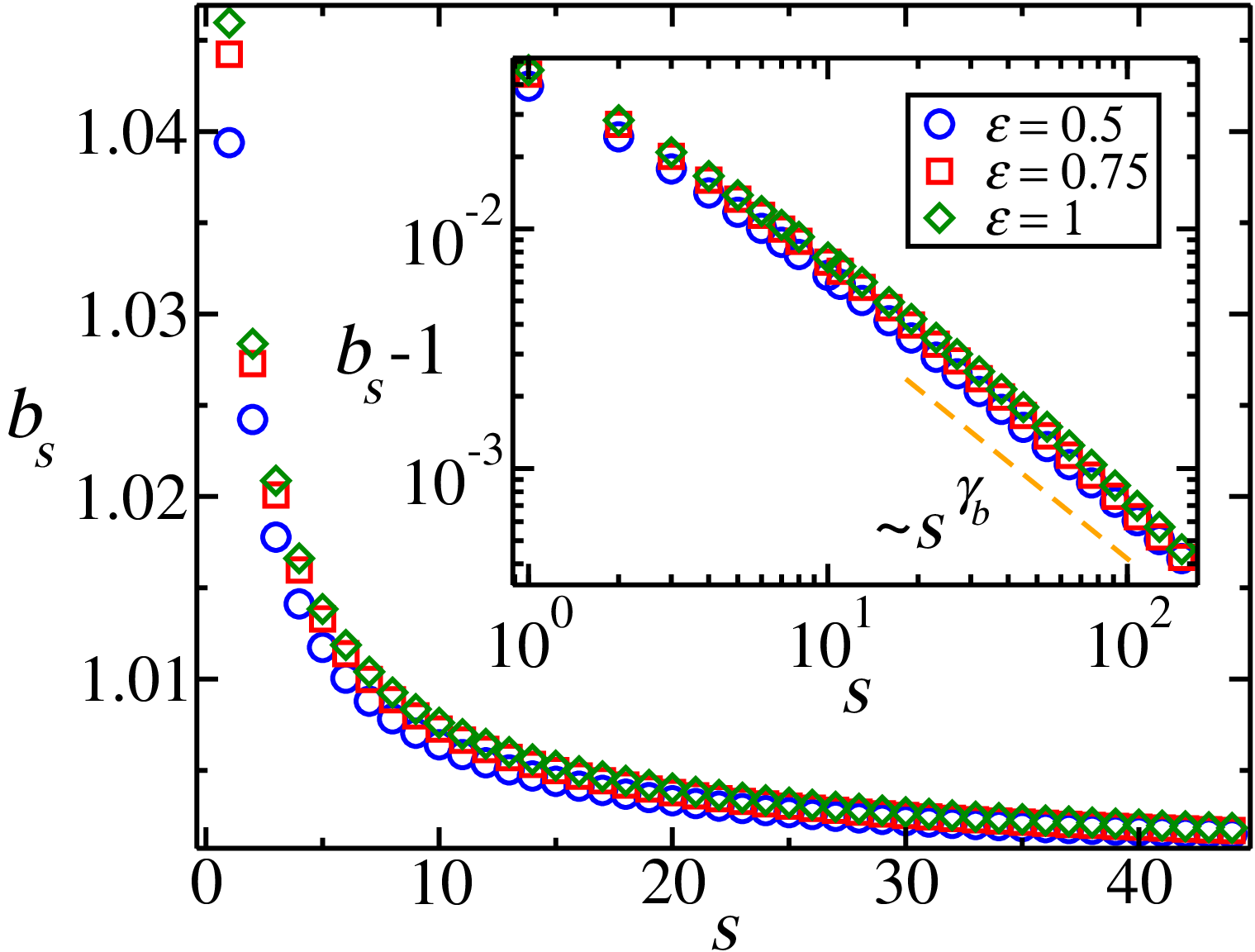}
\caption{{\bf Restricted numerical integration of the KPZ equation.} Plot of the linear coefficient $b_{s}$ as a function of the discretization step $s$  for several values of the restriction $\varepsilon$. The parameters chosen for the integration are: $\lambda=7.746$, $\nu=0.5$, $D=0.0312$, $\Delta t=0.1$, $\Delta x=1$, and $L=2000$. The semi-log inset plot shows the power-law behavior of $b_{s}-1$ as a function of $s$. The measured value of $\gamma_{s}$ is presented in  Table~\ref{tabledataint}.}\label{Intplot}
\end{figure}

\begin{table}[h]
\centering
\begin{tabular}{|c||c|c|c|c|}
\hline
    $\varepsilon$  &  $\gamma_b$ $(\delta \gamma_b)$ & $b_f$  $(\delta b_f)$ & $\Lambda$ $(\delta \Lambda)$ & $\Lambda/b_f $ \\ \hline
  $0.5$               & $-1.023(3)$ & $0.992(9)$ & $7.682(8)$ & $7.744(78)$\\ \hline
  $0.75$              & $-1.105(7)$ & $1.033(1)$ & $8.006(10)$ & $7.746(18)$ \\ \hline
  $1$              & $-1.104(13)$ & $1.042(4)$ & $8.078(4)$ & $7.745(34)$ \\ \hline 
\end{tabular}
\caption{{\bf Restricted integration of the KPZ equation in d=1.} The integration parameters chosen are $\lambda=7.746$, $\nu=0.5$, $D=0.0312$, $\Delta t=0.1$, $\Delta x=1$, and $L=2000$. Second column: measured values of the exponent $\gamma_b$ associated with the linear coefficient $b_s$ from equation~\eqref{eq:MSHG-h_m}, which shows a small dependency with the restriction parameter $\varepsilon$. Third column: the linear constant $b_f$ associated with the restriction function. Fourth column: value of $\Lambda$ measured from the tilt method. Fifth column: shows the value of $\Lambda/b_f $, which recover the value of $\lambda$ taken over the integration. The values inside parenthesis are their respective errors. }
\label{tabledataint}
\end{table}

\section{Discussions}
We show that $\gamma_b\approxeq -1$ for RSOS and BD models. Then, for KPZ models in one dimension, through equations~\eqref{bequa}, \eqref{eq:mean-square-grad}, and \eqref{G0}, the MSF of a tilted interface is
\begin{equation}
\bigl\langle[F(m,\nabla h_0)]^2\bigr\rangle=\bigl\langle\abs{\nabla h_m}^2\bigr\rangle - m^2 =\bigl(a_0\pm a_b\,m^2\bigr)\,s^{\gamma_b}\;,
\end{equation}
with the height gradient discretization of equation~\eqref{eq:grad-discr-tilted}. We see that for, at least, one dimensional models $\gamma_b=2(\zeta-1)$. It should be relevant to see if this relationship still holds for higher dimensions.  

\section{Conclusions}
In this work we show that, for discrete growth models with tilted interfaces that belong to the KPZ university class, the mean square fluctuation (MSF) of the height gradient is a linear function of $m^2$, where $m$ is the average slope of the tilted interface. Because this is a main feature of the continuous KPZ equation, we show how, aside from the scaling exponents, this equation and the discrete models are related. Furthermore, the MSF follows a power law as a function of the integration step used, with exponent $2(\zeta-1)$, the same as in other quantities of interest in the saturation (e.g. equation~\eqref{eq:MSHG-vs-L}). This is because the mean square height-gradient (MSHG) increases in $b_s m^2$ when the interface is tilted. We show that, for 1-dimensional discrete KPZ models,  $b_s$  has a power-law dependence with the discretization step $s$, with the same exponent of the MSF. Moreover, the coefficient $b_s$  goes to $1$ when $s$ increases, reaching the result of the continuous KPZ equation. We also show how the nature of the growth rules (restrictive or non-restrictive) affect the way that $b_s$ goes to $1$. These important results emphasize that any study that tries to obtain properties of the discrete growth models should take into account these local dependencies\cite{Krug1990-2,Lam1993,Campos2013}.  

\section*{Acknowledgements}
This work was partially supported by Consejo Nacional de Investigaciones Cient{\'i}ficas y T{\'e}cnicas (CONICET), Argentina, PIP 2014/16 No. 112-201301-00629. R.C.B. thanks M. Sempé for her suggestions on the final manuscript.

\bibliography{KPZ-tilt}
\end{document}